\documentclass[aps,prc,preprint,groupedaddress,showpacs]{revtex4}

\usepackage{graphicx}% Include figure files
\usepackage{dcolumn}% Align table columns on decimal point
\usepackage{bm}% bold math

\begin{document}

%\preprint{}

\title{Measurement of the $\beta - \nu$ Correlation of $^{21}$Na using Shakeoff Electrons}


\author{P. A. Vetter}
%\thanks{}
\affiliation{Lawrence Berkeley National Laboratory, Berkeley, California 94720}

\author{J. R. Abo-Shaeer}
\affiliation{Lawrence Berkeley National Laboratory, Berkeley, California 94720}

\author{S. J. Freedman}
\affiliation{Department of Physics, University of California at Berkeley and
Lawrence Berkeley National Laboratory, Berkeley, California 94720}

\author{R. Maruyama}
\affiliation{Department of Physics, University of California at Berkeley and
Lawrence Berkeley National Laboratory, Berkeley, California 94720}


\date{\today}

\begin{abstract}
  
The \mbox{$\beta - \nu$} correlation coefficient, $a_{\beta\nu}$, is measured in $^{21}$Na by detecting the time-of-flight of the recoil nucleus detected in coincidence with the atomic electrons shaken off in beta decay.  The sample of $^{21}$Na is confined in a magneto-optic trap.  High detection efficiency allows low trap density, which suppresses the photoassociation of molecular sodium, which can cause a large systematic error.  Suppressing the fraction of trapped atoms in the excited state using a dark trap also reduces the photoassociation process, and data taken with this technique are consistent.  The main remaining systematic uncertainties come from the measurement of the position and size of the atom trap, and the subtraction of background.  We find \mbox{$a_{\beta\nu}=0.5502(60)$}, in agreement with the Standard Model prediction of \mbox{$a_{\beta\nu}=0.553(2)$}, and disagreeing with a previous measurement which was susceptible to an error introduced by the presence of  molecular sodium.

\end{abstract}

%\keywords{}

\maketitle


\section{introduction}

Precise measurements of beta-decay kinematic correlations test the Standard Model of the Weak Interaction. The correlation of the momenta of the $\beta$ and the $\nu$ is sensitive to scalar and tensor current interactions which are suggested by some Beyond Standard Model theories \cite{se06,na91,gl98}.  Neutral atom optical traps containing short-lived isotopes are an appealing source for the study of $\beta$ decay correlations, and several experiments have been performed and proposed in such systems \cite{lu94, go05, fe07, wi07}.  Recoiling daughter nuclei (energy $\sim$100\,eV) and $\beta$ particles emerge from the trap volume with negligible scattering and propagate in ultra-high vacuum.  Decays occur essentially at rest (temperature $<$250\,$\mu$K) and are localized in a volume smaller than 1\,mm$^{3}$.  The trapped atoms are isotopically pure, and very little background activity is present when the detection region is restricted to a small zone around the trap.  The source location and spatial distribution can be monitored by observing atomic fluorescence. The polarization state of the nuclei can be manipulated using optical pumping techniques.  

A previous measurement of the beta-neutrino correlation in laser-trapped $^{21}$Na found a value $3.6\sigma$ smaller than that calculated from the ``V minus A" current coupling of the Standard Model \cite{sc04}. In that work, the results suggested a dependence on the number of atoms: at lower trap populations, the measured correlation coefficient seemed larger, and if extrapolated to zero trap population, the result agreed with the calculated value.  This tentative conclusion is addressed here:  repeating the experiment using a more efficient detection system, we find convincing evidence that the previous measurement of $a_{\beta \nu}$ was distorted by events originating from cold, trapped molecules of $^{21}$Na$_{2}$.  A re-measurement of $a_{\beta \nu}$ has been made with particular attention to minimizing the effect of molecular sodium.

\section{The Beta-Neutrino Correlation in the Electroweak Standard Model} 
In a source with no net nuclear polarization or tensor alignment, the \mbox{$\beta - \nu$} correlation can be inferred from the $\beta$ decay rate \cite{ja57,ho74}
\begin{eqnarray} 
\label{eq:diff-decay}
\frac{d^{3}\Gamma}{dE_{e}d\Omega_{e}d\Omega_{\nu}}
\propto F(Z,E_{e})p_{e}E_{e}(E_{0}-E_{e})^{2}
\nonumber\\
\times\bigg(f_{1}(E_{e}) + 
a_{\beta\nu}(E_{e})\frac{\vec p_{e}\cdot\vec p_{\nu}}{E_{e}E_{\nu}} +
b_{\rm Fierz}(E_{e})\frac{m_{e}}{E_{e}}\bigg)
\end{eqnarray}
by detecting the $\beta$ and recoiling nucleus in coincidence.  Here $(E_{e},\vec p_{e})$ and $(E_{\nu},\vec p_{\nu})$ are the $\beta$ and $\nu$ 4-momentum, $E_{0}$ is the $\beta$ decay endpoint energy, $m_{e}$ is the electron mass, and $F(Z,E_{e})$ is the Fermi function from \cite{be69}.  The energy distribution of the recoil nuclei is calculable (see \cite{na68,gl98}) but gives no greater insight into the decay rate dependence, and the calculation is most simply performed in terms of integrals over $\beta$ energy.  In the allowed approximation, $f_{1}$, the \mbox{$\beta - \nu$} correlation coefficient ($a_{\beta\nu}$) and Fierz interference term ($b_{\rm Fierz}$) are independent of $E_{e}$.  Their values are determined by fundamental weak coupling constants $C_{i}$ and $C'_{i}$ $\{i$ = scalar ($S$), vector ($V$), tensor ($T$), and axial-vector ($A$)$\}$, and by the Fermi (Gamow-Teller) nuclear matrix elements, $M_{F}$ ($M_{GT}$) \cite{ja57}.  
For a mixed Fermi/Gamow-Teller beta decay,
\begin{equation}
a_{\beta\nu} =
\bigg(|M_{F}|^{2}(|C_{V}|^{2}+|C'_{V}|^{2}-|C_{S}|^{2}-|C'_{S}|^{2})
- \frac{1}{3}|M_{GT}|^{2}(|C_{A}|^{2}+|C'_{A}|^{2}-|C_{T}|^{2}-|C'_{T}|^{2})\bigg)
\xi^{-1}
\label{eq:a_coeffs}
\end{equation}
where
\begin{equation}
\xi=|M_{F}|^{2}(|C_{V}|^{2}+|C'_{V}|^{2}+|C_{S}|^{2}+|C'_{S}|^{2})
+|M_{GT}|^{2}(|C_{A}|^{2}+|C'_{A}|^{2}+|C_{T}|^{2}+|C'_{T}|^{2}), 
\label{eq:xi}
\end{equation}
and
\begin{equation}
b_{\rm Fierz} =  \pm 2 \sqrt{1-\left(Z\alpha\right)^{2}} \, Re\left[ |M_{F}|^{2} \left(C_{S}C_{V}^{\ast} + C_{S}'C_{V}'^{\ast} \right) 
+ |M_{GT}|^{2} \left( C_{T} C_{A}^{\ast} + C_{T}' C_{A}'^{\ast} \right) \right]   \xi^{-1} .
\label{eq:fierz}
\end{equation}
In the Standard Model (SM), $C_{V}$ and $C_{A}$ are nearly purely real (and we neglect any imaginary component for the rest of our analysis), \mbox{$C_{V}=C'_{V} = 1$}, and \mbox{$C_{A}=C'_{A}\approx -1.27$} (from experiments), and all other coupling constants are zero.  Experimental limits on scalar and tensor couplings predicted by some SM extensions are model dependent (depending on assumptions about the parity and time-reversal properties of new interactions), and not stringent in some cases \cite{se06}.  If present, these couplings would alter $a_{\beta\nu}$ and could cause non-zero Fierz interference terms. 

Several corrections alter the allowed approximation prediction of $a_{\beta \nu}$ and give $E_{e}$ dependence to $f_{1}$, $a_{\beta\nu}$, and $b_{\rm Fierz}$ at the 1\% level.  The ground state $^{21}$Na$(3/2 ^{+})$ decays via two main branches to $^{21}$Ne$(3/2 ^{+}, 0 \mathrm{keV})$ and $^{21}$Ne$(5/2 ^{+}, 350.7 \mathrm{keV})$.  Electron capture contributes a 0.1\% branching ratio.  From isospin symmetry of the mirror decay $^{21}$Na\,(3/2$^{+}$)$\rightarrow^{21}$Ne\,(3/2$^{+}$), \mbox{$M_{F}=1$}.  The Gamow-Teller contribution, $\lambda = |C_{A}M_{GT}|$, can be determined by the ratio of the $ft$ value to the $(0^{+}\rightarrow0^{+})$ superallowed transitions.  The $ft$ value for the ground state transition $(3/2^{+})$ is determined by the half-life $t_{1/2} = 22.49(4)$ s, and the total decay energy   3547.07(21) from \cite{mu04}.  The recoil order corrections were evaluated in Ref.\ \cite{ia06} and include order-$\alpha$ radiative, and isospin symmetry breaking corrections.  We calculate $ft =  4036.7(9.0)$ and \mbox{$a_{\beta\nu}=0.553(2)$} for $^{21}$Na decays to the ground-state, which is substantially the same as in \cite{ia06}, but accounts for the most recent decay energy measurement.  The decay to the ($5/2^{+}, 350.7$ keV) state is pure Gamow-Teller, and so $a_{\beta \nu}(5/2^{+}) = -1/3$ (from Eqs.\ \ref{eq:a_coeffs} and \ref{eq:xi} when $M_{F} = 0$) to a good approximation.   The omission of recoil order and order-$\alpha$ radiative corrections, the excited-state lifetime, and \mbox{$\beta - \gamma$} correlation is inconsequential for the decays to the  ($5/2^{+}, 350.7$ keV) state. The decay branching ratio to this state has been re-measured to be 4.74(4)\%  \cite{ia06}, and 4.85(12)\% \cite{KVI06}, compared to the previously accepted value of 5.02(13)\%.  The new measurements confirm that the branching ratio was not a source of error in our previous measurement of $a_\beta \nu$.

The magnetic moments of the $^{21}$Na \cite{am65} and $^{21}$Ne \cite{ra89} nuclei determine the weak magnetism recoil order term, characterized by $b_{\rm WM} = 82.6$, assuming the conserved vector current hypothesis.  The absence of second-class currents implies that the induced tensor term is zero for an isodoublet transition.   Order-$\alpha$ radiative effects are included in our calculations of the decay phase space according to the prescription in Ref.\ \cite{gl97}, which accounts for hard bremsstrahlung emission.

\section{\mbox{$\beta - \nu$} measurement technique}
\begin{figure}
\includegraphics[width=0.45\textwidth]{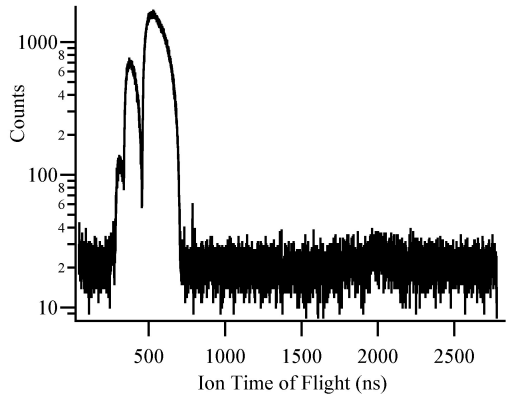}
\caption{\label{fig:wholeTAC} Recoil $^{21}$Ne TOF data.  The time bin width is 0.43\,ns.  Charge states up to +4 are resolved, and neutral $^{21}$Ne is observed as a peak at about 2000 ns.  Molecular $^{21}$Na$_{2}^{+}$ was also detected during this run and is the peak near 800 ns.}
\end{figure}

Much of the experiment is similar to the apparatus and technique described previously in \cite{CSD,sc04,sc04a}.  Sodium-21 is produced via the $^{24}$Mg(p,$\alpha$)$^{21}$Na reaction at the 88" Cyclotron at LBNL.   A beam current of between one and two microamps of 25 MeV protons is incident on a natural abundance magnesium oxide target enclosed in a sealed, heated, ceramic oven.  The $^{21}$Na diffuses from thin disks of the target material (ten disks of $\approx 40$ mg/cm$^{2}$), and emerges from the oven (at about 1000 C) as a neutral atomic beam.  The beam is collimated with 20 mm long by 0.6 mm diameter tubes and with a two-dimensional optical molasses region just downstream from the oven.  The atoms are slowed using a tapered magnetic field and counter-propagating laser beam (Zeeman slower).  The slowed fraction of the atomic beam is loaded into a magneto-optic trap formed by three perpendicular, retroreflected laser beams of about 10 mW/cm$^{2}$ each.   The magneto-optic trap (MOT) is located near the center of the counting chamber containing two microchannel plates (MCPs) and several electrodes to form a focusing electric field of about 1.6 kV/cm in the region of the trap.  The microchannel plates are each about 87 mm away from the trap along a single axis $\left( \vec{x} \right) $.  A sample of $^{21}$Na is accumulated in the magneto-optic trap.  The $\beta$ decay leaves $^{21}$Ne in a variety of charge states through shakeoff and Auger processes \cite{CSD}.  The electric field accelerates the ionized recoil nuclei to one MCP biased at about -7 kV (in the $+\hat{x}$ direction), and the low-energy electrons shaken off by the $^{21}$Ne towards the second MCP biased at about +5.6 kV ($-\hat{x}$ direction).  A trigger from the electron MCP starts a 3\,$\mu$s coincidence window on a time to amplitude converter (TAC), with a stop signal provided by the $^{21}$Ne ions detected by the ion MCP.  The TAC signal is digitized (along with pulse height and rate information) to count the time-of-flight (TOF) spectrum for the recoil ions.  The \mbox{$\beta - \nu$} correlation can be inferred from the TOF spectrum since aligned lepton momenta (caused by the $a_{\beta\nu}$ term in Eq.\ \ref{eq:diff-decay}) result in larger nuclear recoil energies.  An example of the TOF spectrum data is shown in Fig.\ \ref{fig:wholeTAC}, representing about seven hours of data acquisition with traps containing an average of $5\times 10^{4}$ atoms.  Events from neutral $^{21}$Ne are visible as a peak at about 2000 ns.  The detection efficiency for neutral recoils is low, since these recoils are unfocussed by the electric field, and because of the low intrinsic efficiency of the MCP (which has no cesium iodide surface coating as in our previous configuration).  Also visible is a peak at 786 ns caused by autoionized $^{21}$Na$_{2}^{+}$ formed in the MOT (see Section \ref{sec:molecules} below).  During data acquisition, a CCD camera acquires images of the MOT approximately every ten minutes.
The trap population is inferred by measuring the intensity of the fluorescence of the trapped atoms using a calibrated photomultiplier tube in current mode.  

\section{Generating Fit Templates}

\begin{figure}
\includegraphics{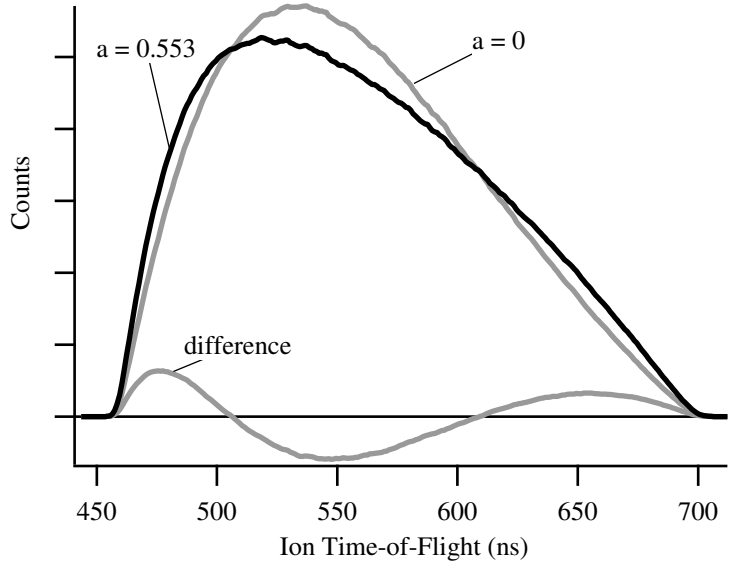}
\caption{\label{fig:MC_TOF} Monte Carlo simulation of the time-of-flight spectra for \mbox{$^{21}$Ne$^{+}$} given
\mbox{$a_{\beta\nu}=0.553$} (calculated value) and \mbox{$a_{\beta\nu}=0$}, and the difference between these two spectra.  There are roughly $4\times 10^{6}$ counts in the spectra.}
\end{figure}

To interpret the time-of-flight spectra as a measurement of $a_{\beta \nu}$, we use Monte-Carlo simulations to generate template TOF curves for the two kinematic terms in Eq.\ \ref{eq:diff-decay} (ignoring $b_{\rm Fierz}$) with $a_{\beta \nu} = 0$ and $a_{\beta \nu} = 0.553$.  The recoil ion TOF data are then fit to a linear combination of these two template spectra to determine the magnitude of $a_{\beta \nu}$.  We construct recoil ion momentum spectra by generating random lepton momenta which are consistent with the phase-space distribution (Eq.\ \ref{eq:diff-decay}), and then obtain the recoil momenta by momentum conservation.  The recoil ions are assigned a random initial position consistent with the spatial distribution of the trapped atom cloud as determined from camera images of the trap. The recoil ion trajectories are calculated using an ion optics program which calculates the electric field $\left( \vec{E}\left(\vec{x}, \vec{y}, \vec{z}\right) \right)$ given the applied electrode voltages and a full three-dimensional description of their geometry (SimIon version 8.0). A Monte-Carlo event is accepted after applying  position- and energy-dependent efficiency functions for the microchannel plate detector.  The Monte-Carlo generated TOF spectra for beta decays to singly charged $^{21}$Ne, calculated with $a_{\beta \nu} = 0$ and $a_{\beta \nu} = 0.553$ are shown in Fig.\ \ref{fig:MC_TOF}.  We generate events for decays both to the $^{21}$Ne ground state and to the $^{21}$Ne($5/2^{+}, 350.7 \, \, \mathrm{kev}$) state, weighted by the branching ratio.   Electron capture events are included with a 0.1\% branching ratio.  The effect of electron capture events in the time-of-flight templates is not resolved in our data analysis to measure $a_{\beta \nu}$.

\subsection{Trap cloud parameters}

Measuring the size and location of the trapped atom cloud with respect to the MCP detectors is crucial for determining $a_{\beta \nu}$.  For each data run, the position and size of the magneto-optic trap were determined from camera images and measured time-of-flight distributions.  The scale of the camera image was determined using calibrating rulers at variable distance from the camera lens.  The MOT position was measured with respect to locating cross hairs (20 micron wires) placed at the entrance and exit optical ports along the viewing axis of the camera.  The cross-hair camera pixel locations were observed at the beginning and end of each data group to correct for any camera movement.  This locates the MOT in a plane perpendicular to the camera viewing axis (the $y'-z$ plane, where $\hat{y}'$ is oriented at 45 degrees with respect to vertical ($+\hat{y}$ and $+\hat{x}$), with an uncertainty of 0.04 mm.  The location of the cross-hair based reference point with respect to the microchannel plate was determined in a calibration using molecular ions' time-of-flight.  The MOT location was varied along the $\hat{y}$ and $\hat{z}$ axes using shimming magnetic fields.  Using an electric field a factor of two lower than is used to detect $\beta$ decay recoils (to avoid strong focusing of the low-energy dimer ions), we measured the TOF of autoionized dimer ions ($^{23}$Na$_{2}^{+}$) as a function of trap position $\vec{y}$ and $\vec{z}$, relative to the cross hair reference point.  For the nearly uniform electric field directed along $\vec{x}$ towards the ion MCP, the TOF has a two-dimensional parabolic dependence on the trap's plane position $\left(\vec{y}, \vec{z}\right)$.  The minimum TOF occurs nearly at the center of the ion MCP in the $y-z$ plane (verified by ion trajectory computation).  The dimer ions' time of flight at this location depends only on the $\vec{x}$ position, i.e.\ the distance to the microchannel plate.  The absolute time of flight was determined from the TAC value with an off-line calibration of the TAC using a precision digital delay generator.   The $T = 0$ offset of the TAC was determined from coincident events in which secondary electrons from the surface of the ion MCP, propagated back to the electron MCP, with a delay from the electron flight time calculated at $\Delta T = -8.9(2)$ ns.  The $\vec{x}$ position of the MOT was obtained using the calculated dimer ions' trajectory.  
The trap location in the $yz$ plane was measured with a position uncertainty of 0.2 mm.  The absolute TOF of the ionized dimers at the location of the minimum TOF determined the $\vec{x}$ position to an uncertainty of 0.01 mm, (accounting for electric field uncertainty).  
The distance from the MOT to the ion MCP was verified by measuring the TOF of $^{21}$Na$_{2}^{+}$ during a beta-decay data run.  In this case, the TOF of the radioactive dimer ions was sensitive mainly to the $\vec{x}$ position, when the $\vec{y}, \vec{z}$ position was known from the camera image.  This measurement gave an $x$ axis position of $86.84(1)$ mm from the ion MCP surface for this run.   The $x$ axis distance was also measured using $\beta$-decay coincidence events of neutral recoils, $^{21}$Ne$^{0}$, finding $87.0(3)$ mm.  In this case, the rising edge of the TOF spectrum depends on the MOT/MCP distance, independent of the electric field, and nearly independent of the value of $a_{\beta\nu}$ and the size of the MOT cloud.  This technique was used only in some of the data runs with resolved $^{21}$Ne$^{0}$ events.  

The spatial distribution of the atoms in the MOT was assumed to be gaussian in the Monte-Carlo procedure.  The full-width at half-maximum (FWHM) of the MOT spatial distribution was determined by a least-squares fit to the camera images. A typical FWHM was 0.6 mm, varying by about 30\% between different data groups, depending on the MOT laser and magnetic field parameters.  The gaussian FWHM was up to 30\% different along different axes of the trap, and a two-dimensional Gaussian fit with an asymmetry parameter was typcially performed.  During data runs, the MOT position remained constant to within 0.2 mm, determined by the centroids of the fits to the camera images.  To analyze each data group, the position $\left( \vec{x}, \vec{y}, \vec{z} \right)$ and FWHM of the trap were inserted into the Monte-Carlo simulations, and the measured potentials applied to the electrodes and MCPs were used to calculate the electric field $\vec{E}\left(\vec{x}, \vec{y}, \vec{z}\right)$ to propagate simulated ion trajectories.  

\section{\label{sec:molecules}Molecular sodium}
As the MOT operates, molecular sodium (Na$_{2}$) is generated via photoassociation during collisions between cold trapped atoms \cite{fa02}.  Cold, ground state molecule formation, using only MOT laser frequencies, has been measured in cesium \cite{fi98}, while a dedicated experiment observed trapping of triplet ground state Cs$_{2}$ by the quadrupole MOT magnetic field gradient acting on the molecular magnetic moment \cite{va02}.  In sodium, the second molecular excited state manifold (asymptotically equivalent at large interatomic separation to the state with both partners in the $3\,^{2}P_{3/2}$ state) is autoionizing.  This creates a low-energy electron and a Na$_{2}^{+}$ ion with very low initial momentum.  We detect molecular sodium as coincidence events in the MCP pair with a strongly peaked TOF corresponding to singly ionized species with twice the mass of atomic sodium, originating from the trap location.   
In our previous measurement, Ref.\ \cite{sc04}, the result for $a_{\beta\nu}$ depended on the number of atoms held in the MOT, and this dependence was likely caused by counting beta decay events in which the detected recoil nucleus scattered from a molecular partner.  

For beta decay occuring in $^{21}$Na$_{2}$, the recoil nucleus is created near the molecular partner.  The scattering potential for $^{21}$Ne-$^{21}$Na is not known, and it would depend on the charge state of the $^{21}$Ne recoil.  A simple model of the scattering of the $^{21}$Ne recoil nuclei by $^{21}$Na suggests a very strong perturbation of the measured value of $a_{\beta\nu}$.  Scattering of the recoil nuclear momentum will tend to randomize the momentum direction, and the momentum would be shared with the molecular partner, distorting the initial momentum distribution of the $\beta$ decay recoil nuclei.  
The scattering process was modeled using a procedure similar to Ref.\ \cite{vo03},  assuming a simple Lennard-Jones interatomic potential for the recoil $^{21}$Ne with $^{21}$Na.  Monte-Carlo generated recoil momenta were generated at a $^{23}$Na-$^{23}$Na binding distance $r_{0} = 0.28$ nm \cite{ma93}, randomly directed with respect to the scattering potential, and then propagated in the scattering potential to determine final momenta.  These were propagated in the ion optics program to create a TOF spectrum for scattered recoil nuclei.  The simulated, perturbed recoil TOF spectra were fit with simulated, unscattered template TOF spectra to determine the approximate size of the error introduced in a measurement of $a_{\beta \nu}$.  For both the experimental configuration of Ref.\ \cite{sc04} (in which the $\beta$ was detected) and the current configuration (triggered by shakeoff electrons) the value of $a_{\beta \nu}$ is reduced to 0.6 times the nominal value used in the simulations.  The reduction factor depends on the simulated Lennard-Jones parameters, but for a range of realistic values (using the $^{23}$Na - $^{23}$Na parameters in \cite{ma93}), $a_{\beta \nu}$ is reduced by about 0.4 to 0.7.  

The autoionization process in our MOT sample is caused primarily by absorption of laser photons to form singly-excited, short-range molecular states, which then absorb a second laser photon to excite to short-range, autoionizing states.  This process is known as photoassociative photoionization (PAPI), and was shown in Ref.\ \cite{am03} to be the dominant ionization process in experiments using a single laser frequency tuned below the atomic resonance.  The formation of cold, ground state molecules requires the same intermediate step -- the population of short-range singly excited molecular states, which then decay via spontaneous or stimulated emission to bound ground states.  The observation of autoionized dimer molecules implies a population of cold, ground state molecular sodium.  Cold, ground state sodium molecules with a net magnetic moment (e.g.\ in the $^{3}\Sigma_{u}^{+}$ state) can be confined in the magnetic trap formed by the MOT's magnetic field gradient of 10 G/cm, implying a cold, trapped, molecular population.  It is difficult to determine the absolute fraction of trapped $^{21}$Na which is bound as $^{21}$Na$_{2}$ by measuring the rate of observed ionized dimers.  In principle, this would help to determine the size of the systematic error introduced by recoil scattering.  But this would require knowing both the spontaneous emission and autoionization probabilities from both ground and singly excited molecular states.  It is also difficult to know the distribution of  molecular states involved in the photoassociation and autoionization, since the photoassociation occurs with only the MOT lasers, rather than a separate, frequency-tunable laser which probes the molecular spectroscopy.  We have measured the rate of autoionized $^{23}$Na$_{2}$ dimers as a function of the trapped atomic population:  this measures the relative cold molecular population of the MOT as a function of atomic population, since the formation rate of autoionized dimers must be related to the formation rate of cold, ground state, trapped dimers.  We find a strong dependence of the dimer ion rate {\em per trapped atom} (the relevant scaling parameter for perturbing a measurement of $a_{\beta \nu}$) on the population of the MOT, shown in Fig.\ \ref{fig:dimer_rate}.  

\section{Data Analysis}

The total data set consists of $3.6\times 10^{6}$ coincident electron-ion events taken during eight separate runs with a total data acquisition time of about 66 hours.  Steady-state trap populations during these runs ranged from 5000 to 83000 atoms, and the statistically weighted mean number of atoms in the trap for the entire data set was $5\times 10^{4}$. Each run was analyzed separately, generating new Monte-Carlo templates with different input parameters for trap size and location, and different electrode potentials.  To analyze the data, we apply the TAC time calibration (which corrects for a roughly $2 \times 10^{-4}$ differential non-linearity of the TAC and ADC conversion) and add the 4 ns shakeoff electron flight time (with a time spread of 0.2 ns caused mainly by the size of the cloud) to the raw TAC data to generate a true ion TOF spectrum.  The amplitude of the background in the TOF spectrum is parametrized using measured background data (taken with one MOT beam blocked to prevent the formation of the trapped population, but still allowing a slowed radioactive atomic beam  into the detection volume).  Since the TOF is determined from the first ion MCP trigger to arrive after the electron MCP trigger, lost true and accidental coincidences must be corrected using the average ion and electron MCP trigger rates as described in \cite{lu97}.  This requires a slight correction to the TOF spectrum, amounting to roughly 0.3\% of the counts in the $^{21}$Ne$^{+1}$ peak.  The amplitude of the corrected background spectrum is determined by fitting the trap-on beta decay TOF spectrum in regions away from the recoil ion and molecular peaks.  The amplitude of the background term is then fixed in the fits to the recoil-ion TOF spectrum. Another issue in subtracting a background TOF spectrum is the contribution from events caused by the coincident detection of a $\beta^{+}$ (or $\gamma$) with the electron MCP and a positively charged recoil $^{21}$Ne ion.  Such events were detected by biasing the electron MCP and its collimator electrode to reject shakeoff electrons with less than 500 eV originating from the MOT location.  We estimate the intrinsic detection efficiency for $\beta^{+}$ to be roughly 3\% of the efficiency for the shakeoff electrons (accelerated to about $1.3$ keV) normally used as the trigger.  The $\beta^{+}$ (or $\gamma$) triggered events select a direction for the $\beta^{+}$ momentum, favoring recoil ions with a preferred momentum direction towards the ion MCP, which would give a distorted measurement of $a_{\beta \nu}$ in the shakeoff-triggered TOF spectrum.  It is difficult to accurately calculate the contribution to the measured TOF spectrum from these $\beta^{+}$-triggered events, since the detection efficiency of the electron MCP as a function of  $\beta^{+}$ energy and number of incident shakeoff electrons is not precisely known.  The $\beta^{+}$-triggered events should have a slightly enhanced coincidence detection probability since the electron MCP is struck by both a $\beta^{+}$ and the shakeoff electrons given off in the decay.  We correct the measured $\beta^{+}$-triggered spectrum for acquisition time, the slightly different accelerating potential for the ions, and the estimated trigger efficiency enhancement and subtract it from the recoil ion coincidence TOF spectrum before fitting.  Completely neglecting the subtraction of the $\beta$-triggered component to the TOF spectrum changes the fitted value of $a_{\beta\nu}$ by only 0.5\%.

\begin{figure}
\includegraphics[width=0.45\textwidth]{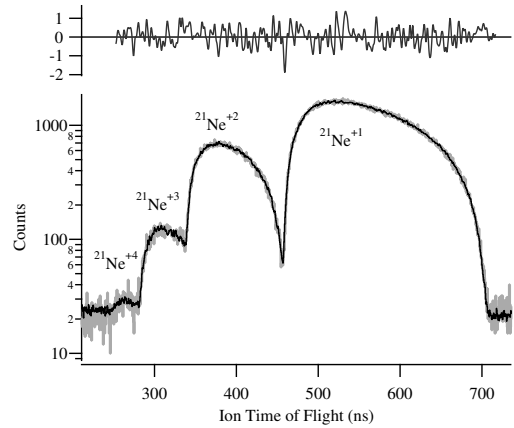}
\caption{\label{fig:TOF_and_Fit} Recoil $^{21}$Ne spectra fit with the MC simulation, with bin width 0.43\,ns.  Residuals are
shown with 4.3\,ns bins for clarity.}
\end{figure}

In fitting the data by chi-squared minimization, the following parameters are variable: the amplitude of each charge state (+1 to +4), an overall time axis shift ($T=0$ offset), and the ratio of the two Monte-Carlo template curves which account for the isotropic term and beta-neurtino momentum correlation term in the beta-decay phase space.  The charge states branching ratios are known in principle (measured in \cite{CSD}), but since the absolute detection efficiency is unknown for the electron MCP detector as a function of the number of shakeoff electrons impinging on it, the strength of each charge state is effectively arbitrary.  Data and fit for one ten-hour data run are shown in Fig.\ \ref{fig:TOF_and_Fit}.  The residuals have been normalized by the statistical uncertainty for each point and binned.

The $\vec{x}$ location of the MOT was verified by fitting the data using several template TOF spectra, generated for a range of $\vec{x}$ positions.  The minimum $\chi^{2}$ occurred with a MOT-MCP distance $D =86.6(2)$ mm in Monte-Carlo simulation  (compared to $D = 87.0(3)$ (based on the $^{21}$Ne$^{0}$ TOF) and $D = 86.84(1)$ (based on the dimer ions and optical reference).

\subsection{Internal Conversion}
A small correction is necessary to the measured value of $a_{\beta \nu}$ because internal conversion of the excited-state $(5/2^{+}, 350.7$ kev) causes the excited state contribution in each charge state of $^{21}$Ne to deviate from the $\beta$ decay branching ratio.  Internal conversion results in an inner shell vacancy subsequent electron loss, causing higher charge states for excited state decays, or effectively larger decay branching ratio to the higher charge states in the data.  Using the same process described in \cite{sc04}, this gives a correction $\Delta a_{\beta \nu} = +0.0003(1)$.  

\subsection{Ionization Dependences}
	Since the \mbox{$\beta - \nu$} correlation is measured only for daughter $^{21}$Ne that have lost $\geq$2 electrons, the ionization process could lead to systematic effects.  As discussed in Ref.~\cite{CSD}, the ratios  
$^{21}$Ne$^{+2}$\,:\,$^{21}$Ne$^{+}$ and $^{21}$Ne$^{+3}$\,:\,$^{21}$Ne$^{+}$ show no indication the $\beta^{+}$ or recoil-ion energy influences ionization.  A rough calculation indicates that nuclear recoil should increase ionization for the fastest recoils by \mbox{0.70(35)\%} \cite{CSD} and we apply a correction of \mbox{$-0.0033(17)$} to $a_{\beta\nu}$.

\subsection{Polarization and Alignment}

The nuclear polarization and alignment of the trapped sodium nuclei have been determined to be $< 0.2\%$ for very broad ranges of trapping parameters by optical rotation spectroscopy on trapped $^{23}$Na atoms \cite{sc04}.  No correction for contributions from polarized nucleus correlations are necessary, although a small uncertainty is applied to the final result for the uncertainty in the measurement of the polarization state of the atoms.  
	
\subsection{Molecular sodium}
\begin{figure}
\includegraphics{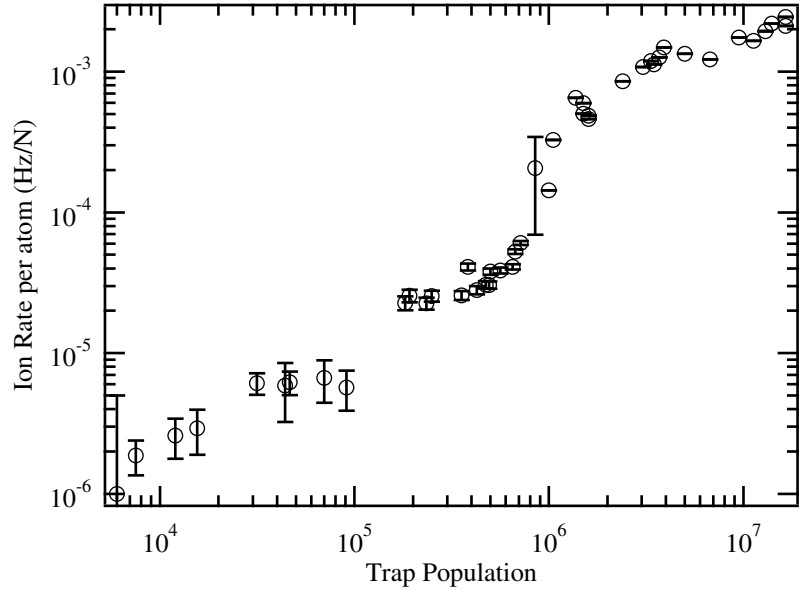}
\caption{\label{fig:dimer_rate} Rate of detected $^{23}$Na$_{2}^{+}$ per trapped atom as a function of the trap atomic population.  The trap population is measured by the MOT's fluorescence.}
\end{figure}

\begin{figure}
\includegraphics{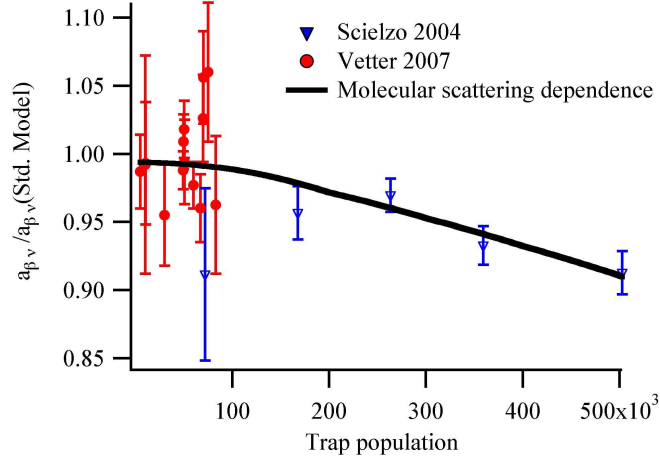}
\caption{\label{fig:a_slope} (Color) $a_{\beta\nu}$ at different trap populations.  Data from \cite{sc04} are also included.  The solid line is a fit to the expected dependence of perturbation from a population of trapped $^{21}$Na$_{2}$, causing scattering of the recoil momenta.}
\end{figure}

To address the problem caused by recoil scattering from molecular partners, data were acquired with a range of trap populations.  Figure \ref{fig:dimer_rate} suggests that there should be a dependence of $a_{\beta \nu}$ with trap population, as the molecular population fraction per trapped atom changes.  With a low trap population, according to Fig.\ \ref{fig:dimer_rate}, we expect a much lower fraction of detected beta-decay events will be from molecular $^{21}$Na$_{2}$. Data were also acquired using a dark MOT technique, in which the usual trap repumping laser is not incident on the trapped atoms, except at the edges of the trapping region \cite{ke93}.  This optically pumps the atoms to $F_{g} = 1$ where they no longer interact with the trapping lasers.  The average $3\, ^{2}P_{3/2}$ excited state population is greatly reduced, inhibiting the photoassociation process (which requires collisions involving excited state atoms).  With the dark MOT technique, the detected rate of  autoionized dimers is reduced by a factor of about 700 relative to the bright MOT.  The measured value of $a_{\beta \nu}$ for dark MOT data (expressed as a fraction of the ``Standard Model" value $a_{\rm SM} = 0.553$) is $a = 1.004(17)$.  This value agrees with $a = 0.994(7)$, the mean of the data measured with bright MOTs.  The dark MOT data was taken with a trap population of 50000 atoms.  In Fig.\ \ref{fig:a_slope}, we show the current data set and the averaged data from \cite{sc04}.  The perturbation on $a_{\beta \nu}$ from molecular recoil scattering can be estimated by fitting these data to a curve derived from the measured rate of molecular ions (per atom) as a function of trap population, Fig.\ \ref{fig:dimer_rate}.  This fit gives a negligible correction (0.05\%) to the current data set for an extrapolation to zero trap population, while also supporting the plausibility of such an extrapolation for the data measured in \cite{sc04}.  

\subsection{Results}
Averaging the data from the present data set (weighted by the statistical uncertainty for each run) gives $a_{\beta \nu} = 0.5498(38)$.  Several corrections are applied to this value to account for instrument effects and biases caused by the shakeoff ionization process.  These are summarized in Table \ref{ta:systematics}, along with the significant sources of systematic uncertainty.  The efficiency of the ion MCP varies as a function of ion impact location.  This was measured by examining the pulse height spectra of the ion MCP for events from $^{23}$Na$_{2}^{+}$ as a function of ion impact location $\left( \vec{y}, \vec{z}\right)$ on the MCP.   The effective detection diameter of the ion MCP is uncertain in this procedure, and the manufacturer's specification of the detector's active diameter is used to estimate this uncertainty.  The effect of a slight decrease in efficiency at the center of the plate (0.45\%) was modeled using the Monte-Carlo generating procedure to create distorted TOF spectra, and then fitting this simulated data with undistorted simulated data.  A similar technique was applied to study the effect of uncertainty in the electric field caused by uncertainty in measurements of electrode potentials and locations in the ion optics calculation.  The uncertainties in Table \ref{ta:systematics} are summed in quadrature, while the corrections are added linearly.  The result is $a_{\beta \nu} = 0.5502 (38) (46)$, where the first uncertainty is statistical and the second systematic.  Again, this is to be compared with the calculated value $a_{\beta \nu} = 0.553(2)$, which assumes Standard Model current couplings, and which has an uncertainty limited by measurements of the decay parameters (half-life, $Q$ value, and branching ratios).   

\subsection{Conclusions}
\begin{figure}
\includegraphics[width=3.4in]{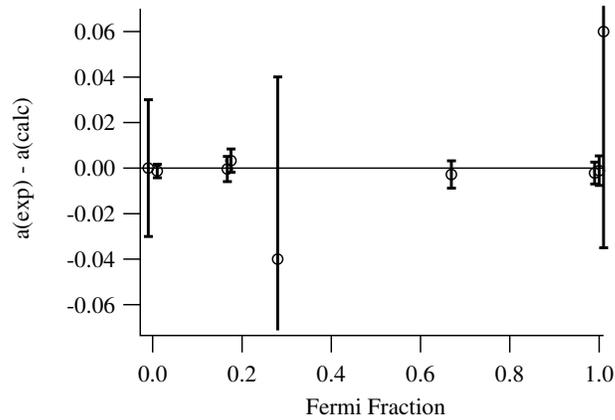}
\caption{\label{fig:a_vs_ferminess} Deviations of measurements of $a_{\beta \nu}$ from calculated values plotted as a function of the Fermi fraction of the $\beta$ decay.  From left to right, the systems are $^{23}$Ne \cite{al63}, $^{6}$He \cite{jo63}, neutron \cite{st78,by02}, $^{19}$Ne \cite{al63}, $^{21}$Na (this work), $^{38m}$K \cite{go05}, $^{32}$Ar \cite{ad99}, and $^{18}$Ne \cite{eg97}.}
\end{figure}

\begin{figure}
\includegraphics[width=3.4in]{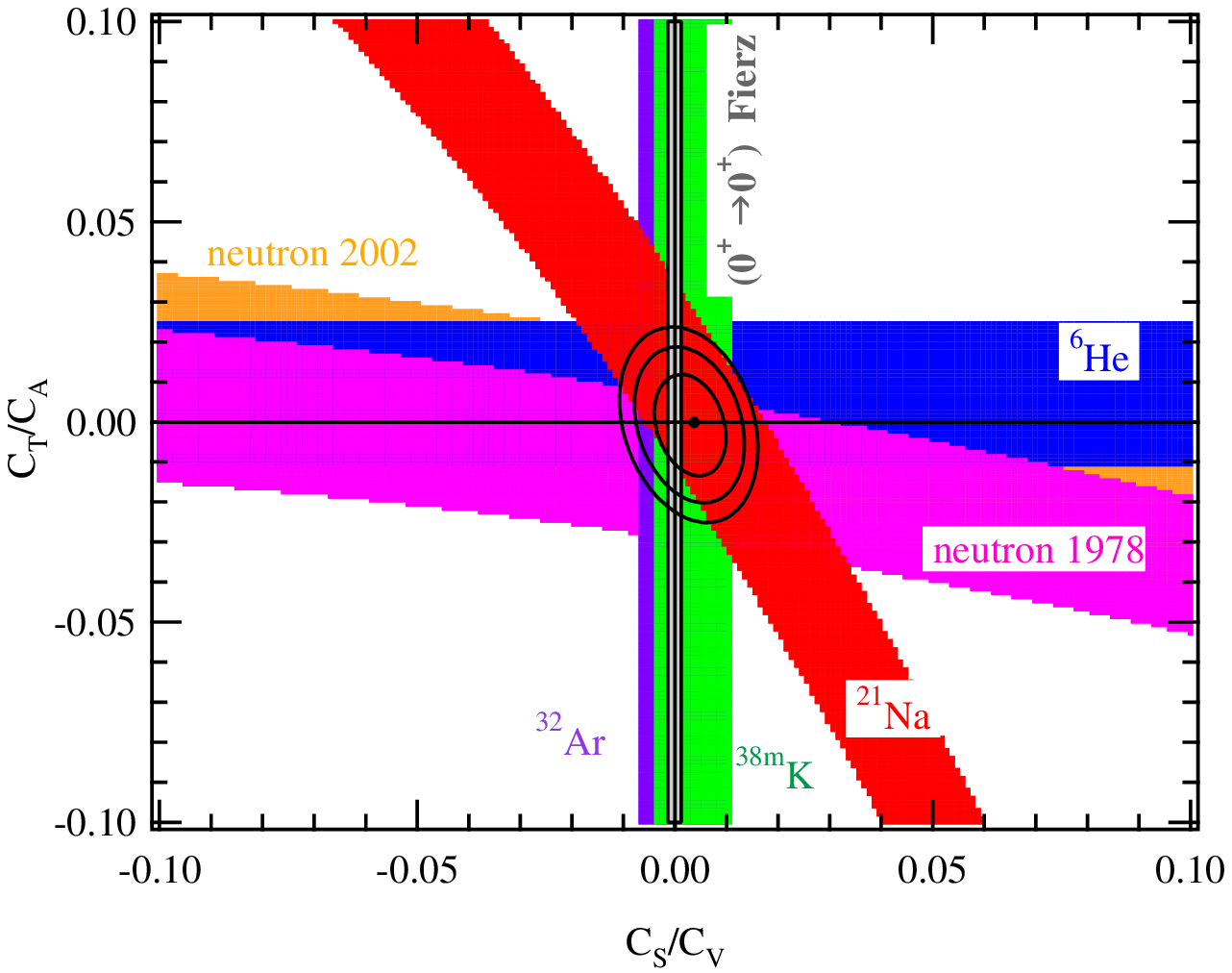}
\caption{\label{fig:exclusioncs} (Color) Allowed regions for scalar and tensor coupling constants (assuming normal helicity) for precise measurements of $a_{\beta\nu}$.  The one, two, and three standard deviation contours are shown at the center.  The vertical grey band is the allowed region for scalar coupling constrained by the limit on the Fierz term from the agreement of the ${\cal F}t$ values of superallowed $(0^{+} \rightarrow 0^{+})$ Fermi decays \cite{ha05}.  
The systems are $^{6}$He \cite{jo63} (blue), neutron \cite{st78} (pink) and \cite{by02} (orange), $^{21}$Na (this work, red), $^{38m}$K \cite{go05} (green), and $^{32}$Ar \cite{ad99} (violet).  The relatively low-precision measurement for $^{23}$Ne \cite{al63} is not shown as a limit band, but is used to calculate the combined allowed region contours.}
\end{figure}

In Fig.\ \ref{fig:a_vs_ferminess} we plot precise measurements of the $\beta - \nu$ correlation for different nuclear systems as a function of the Fermi fraction of the transition, $  F = \left( 1 + \left| C_{A}M_{GT}/C_{V} M_{F} \right| ^{2} \right)^{-1}$.  We show the residual to the theoretical dependence (the measured $a_{\beta \nu}$ minus the ``V minus A" calculated value $a_{\beta\nu}(\mathrm{Std. Model})$).  A plot of $a_{\beta \nu}$ from Eq.\ \ref{eq:a_coeffs} versus $F$ would be a straight line with offset $-1/3$ and slope $+4/3$.  Our result is seen to have precision comparable to the measurements in $^{6}$He, $^{38m}$K, $^{32}$Ar, and the neutron.  In these experiments, the beta neutrino correlation is measured essentially from the recoil energy spectrum, meaning that the observable is 
\begin{equation}
\tilde{a} = \frac{a}{1 + b_{\rm Fierz} \frac{m_{e}}{\left\langle E_{\beta} \right\rangle} } , 
\end{equation}
with $a$ and $b_{\rm Fierz}$ as in Eqs. \ref{eq:a_coeffs} and \ref{eq:fierz}, and $\left\langle E_{\beta} \right\rangle$ the mean energy of the $\beta^{\pm}$.  Combining the results of several precise measurements of the beta-neutrino correlation $\tilde{a}$ produces 
 a limit on the existence of non-Standard Model scalar and tensor currents, shown in Fig.\ \ref{fig:exclusioncs}.  To generate this exclusion plot, we assume "normal helicity" $C_{S} = C_{S}'$ and $C_{T} = C_{T}'$, and we assume that $Im \left( C_{S} \right) =Im \left( C_{T} \right) = 0$ (i.e. time reversal invariance).  Each system, by virtue of the different contriubtions of Fermi and Gamow-Teller transition strengths, and different mean beta energies for the Fierz term contribution, yields a different sensitivity to possible scalar and tensor contributions.  The one, two, and three standard deviation contours are derived from the combined constraints offered by the different measurements of $\tilde{a}$.  For comparison, we also include on the graph (as a vertical grey bar very near $C_{S} = 0$) the allowed region (at one standard deviation) of scalar coupling constants consistent with the analysis of superallowed Fermi decays \cite{ha05}.  In these systems, the agreement of the ${\cal F} t$ values of many beta decays despite different Q values provides a constraint to the size of the Fierz term in the decay phase space suggested in Eq.\ \ref{eq:diff-decay}.  
 
Looking forward, measurements of the beta-neutrino correlation coefficient using laer-trapped atoms could likely be improved to near the 0.1\% level of precision, which would offer even more potent limits on Beyond Standard Model physics.  In both our experiment and that in Ref. \cite{go05}, the uncertainties were limited by detector response calibration, electric field calibrations, and measurement of the trapped atom cloud distribution and location (i.e. the source function uncertainty).  The uncertainty generated by the momentum dependence of the recoil ionization probability could be reduced by a more sophisticated treatment of the problem than the semi-empirical calculation given in Ref.\ \cite{CSD}.  The uncertainty in the predicted value for $a_{\beta \nu}$ (0.4\%,  dominated by the uncertainties in the half-life and decay branching ratio) would then be a limiting factor in interpreting a result as a constraint on Beyond Standard Model couplings.

\begin{table}
\caption{\label{ta:systematics}Corrections and systematic uncertainties.}
\begin{ruledtabular}
\begin{tabular}{lcc}
Source & Correction (\%) & Uncertainty (\%) \\
\hline
Recoil ionization & $-$0.4 & 0.2 \\
Internal conversion & 0.54 & 0.13 \\
Polarization and alignment & 0 & 0.1 \\
$\beta$ triggered background & $-$0.5 & 0.4 \\
$\mathcal{E}_{MCP}$ & 0.25 & 0.15 \\
Electric field and simulation & 0 & 0.27 \\
MCP diameter uncertainty &  & 0.16 \\
Trap FWHM & & 0.5 \\
Trap position ($\vec{x}$) & & 0.2 \\
Trap position ($\vec{y}, \vec{z}$) & & +0.3 \\
\hline
Total & $-$0.11 & 0.85 \\
\end{tabular}
\end{ruledtabular}
\end{table}

We appreciate the assistance of the technical staff and operators at the 88-Inch Cyclotron.  We thank N.D.\ Scielzo for early data acquisition and discussions during analysis. We thank E. Oelker and L. Martinez for assistance with calibration data.  This work was supported by the Director, Office of Science, Office of Nuclear Physics, U.S.\ Department of Energy under Contract No.\ DE-AC02-05CH11231f.
% \end{acknowledgments}

%\bibliography{beta-nushake-arxiv.bib}

\begin{thebibliography}{35}
\expandafter\ifx\csname natexlab\endcsname\relax\def\natexlab#1{#1}\fi
\expandafter\ifx\csname bibnamefont\endcsname\relax
  \def\bibnamefont#1{#1}\fi
\expandafter\ifx\csname bibfnamefont\endcsname\relax
  \def\bibfnamefont#1{#1}\fi
\expandafter\ifx\csname citenamefont\endcsname\relax
  \def\citenamefont#1{#1}\fi
\expandafter\ifx\csname url\endcsname\relax
  \def\url#1{\texttt{#1}}\fi
\expandafter\ifx\csname urlprefix\endcsname\relax\def\urlprefix{URL }\fi
\providecommand{\bibinfo}[2]{#2}
\providecommand{\eprint}[2][]{\url{#2}}

\bibitem[{\citenamefont{Severijns et~al.}(2006)\citenamefont{Severijns, Beck,
  and Naviliat-Cuncic}}]{se06}
\bibinfo{author}{\bibfnamefont{N.}~\bibnamefont{Severijns}},
  \bibinfo{author}{\bibfnamefont{M.}~\bibnamefont{Beck}}, \bibnamefont{and}
  \bibinfo{author}{\bibfnamefont{O.}~\bibnamefont{Naviliat-Cuncic}},
  \bibinfo{journal}{Rev.\ Mod.\ Phys.} \textbf{\bibinfo{volume}{78}},
  \bibinfo{pages}{991} (\bibinfo{year}{2006}).

\bibitem[{\citenamefont{Naviliat-Cuncic
  et~al.}(1991)\citenamefont{Naviliat-Cuncic, Girard, Deutsch, and
  Severijns}}]{na91}
\bibinfo{author}{\bibfnamefont{O.}~\bibnamefont{Naviliat-Cuncic}},
  \bibinfo{author}{\bibfnamefont{T.~A.} \bibnamefont{Girard}},
  \bibinfo{author}{\bibfnamefont{J.}~\bibnamefont{Deutsch}}, \bibnamefont{and}
  \bibinfo{author}{\bibfnamefont{N.}~\bibnamefont{Severijns}},
  \bibinfo{journal}{J.\ Phys.\ G} \textbf{\bibinfo{volume}{17}},
  \bibinfo{pages}{919} (\bibinfo{year}{1991}).

\bibitem[{\citenamefont{Gl$\ddot{\mathrm{u}}$ck}(1998)}]{gl98}
\bibinfo{author}{\bibfnamefont{F.}~\bibnamefont{Gl$\ddot{\mathrm{u}}$ck}},
  \bibinfo{journal}{Nucl.\ Phys.\ A} \textbf{\bibinfo{volume}{628}},
  \bibinfo{pages}{493} (\bibinfo{year}{1998}).

\bibitem[{\citenamefont{Lu et~al.}(1994)\citenamefont{Lu, Bowers, Freedman,
  Fujikawa, Mortara, Shang, Coulter, and Young}}]{lu94}
\bibinfo{author}{\bibfnamefont{Z.-T.} \bibnamefont{Lu}},
  \bibinfo{author}{\bibfnamefont{C.~J.} \bibnamefont{Bowers}},
  \bibinfo{author}{\bibfnamefont{S.~J.} \bibnamefont{Freedman}},
  \bibinfo{author}{\bibfnamefont{B.~K.} \bibnamefont{Fujikawa}},
  \bibinfo{author}{\bibfnamefont{J.~L.} \bibnamefont{Mortara}},
  \bibinfo{author}{\bibfnamefont{S.-Q.} \bibnamefont{Shang}},
  \bibinfo{author}{\bibfnamefont{K.~P.} \bibnamefont{Coulter}},
  \bibnamefont{and} \bibinfo{author}{\bibfnamefont{L.}~\bibnamefont{Young}},
  \bibinfo{journal}{Phys.\ Rev.\ Lett.} \textbf{\bibinfo{volume}{72}},
  \bibinfo{pages}{3791} (\bibinfo{year}{1994}).

\bibitem[{\citenamefont{Gorelov et~al.}(2005)}]{go05}
\bibinfo{author}{\bibfnamefont{A.}~\bibnamefont{Gorelov}} \bibnamefont{et~al.},
  \bibinfo{journal}{Phys.\ Rev.\ Lett.} \textbf{\bibinfo{volume}{94}},
  \bibinfo{pages}{142501} (\bibinfo{year}{2005}).

\bibitem[{\citenamefont{Feldbaum et~al.}(2007)\citenamefont{Feldbaum, Wang,
  Weinstein, Vieira, and Zhao}}]{fe07}
\bibinfo{author}{\bibfnamefont{D.}~\bibnamefont{Feldbaum}},
  \bibinfo{author}{\bibfnamefont{H.}~\bibnamefont{Wang}},
  \bibinfo{author}{\bibfnamefont{J.}~\bibnamefont{Weinstein}},
  \bibinfo{author}{\bibfnamefont{D.}~\bibnamefont{Vieira}}, \bibnamefont{and}
  \bibinfo{author}{\bibfnamefont{X.}~\bibnamefont{Zhao}},
  \bibinfo{journal}{Phys.\ Rev. A} \textbf{\bibinfo{volume}{76}},
  \bibinfo{pages}{051402(R)} (\bibinfo{year}{2007}).

\bibitem[{\citenamefont{Wilschut et~al.}(2007)}]{wi07}
\bibinfo{author}{\bibfnamefont{H.}~\bibnamefont{Wilschut}}
  \bibnamefont{et~al.}, \bibinfo{journal}{Hyperfine Interactions}
  \textbf{\bibinfo{volume}{174}}, \bibinfo{pages}{97} (\bibinfo{year}{2007}).

\bibitem[{\citenamefont{Scielzo
  et~al.}(2004{\natexlab{a}})\citenamefont{Scielzo, Freedman, Fujikawa, and
  Vetter}}]{sc04}
\bibinfo{author}{\bibfnamefont{N.~D.} \bibnamefont{Scielzo}},
  \bibinfo{author}{\bibfnamefont{S.~J.} \bibnamefont{Freedman}},
  \bibinfo{author}{\bibfnamefont{B.~K.} \bibnamefont{Fujikawa}},
  \bibnamefont{and} \bibinfo{author}{\bibfnamefont{P.~A.}
  \bibnamefont{Vetter}}, \bibinfo{journal}{Phys.\ Rev.\ Lett.}
  \textbf{\bibinfo{volume}{93}}, \bibinfo{pages}{102501}
  (\bibinfo{year}{2004}{\natexlab{a}}).

\bibitem[{\citenamefont{Jackson et~al.}(1957)\citenamefont{Jackson, Treiman,
  and Wyld}}]{ja57}
\bibinfo{author}{\bibfnamefont{J.~D.} \bibnamefont{Jackson}},
  \bibinfo{author}{\bibfnamefont{S.~B.} \bibnamefont{Treiman}},
  \bibnamefont{and} \bibinfo{author}{\bibfnamefont{H.~W.} \bibnamefont{Wyld}},
  \bibinfo{journal}{Phys.\ Rev.} \textbf{\bibinfo{volume}{106}},
  \bibinfo{pages}{517} (\bibinfo{year}{1957}).

\bibitem[{\citenamefont{Holstein}(1974)}]{ho74}
\bibinfo{author}{\bibfnamefont{B.~R.} \bibnamefont{Holstein}},
  \bibinfo{journal}{Rev.\ Mod.\ Phys.} \textbf{\bibinfo{volume}{46}},
  \bibinfo{pages}{789} (\bibinfo{year}{1974}).

\bibitem[{\citenamefont{Behrens and J$\ddot{\mathrm{a}}$necke}(1969)}]{be69}
\bibinfo{author}{\bibfnamefont{H.}~\bibnamefont{Behrens}} \bibnamefont{and}
  \bibinfo{author}{\bibfnamefont{J.}~\bibnamefont{J$\ddot{\mathrm{a}}$necke}},
  \emph{\bibinfo{title}{Numerical Tables for Beta-Decay and Electron Capture}}
  (\bibinfo{publisher}{Springer-Verlag}, \bibinfo{address}{New York},
  \bibinfo{year}{1969}).

\bibitem[{\citenamefont{Nachtmann}(1968)}]{na68}
\bibinfo{author}{\bibfnamefont{O.}~\bibnamefont{Nachtmann}},
  \bibinfo{journal}{Z.\ Phys.} \textbf{\bibinfo{volume}{215}},
  \bibinfo{pages}{505} (\bibinfo{year}{1968}).

\bibitem[{\citenamefont{Mukherjee et~al.}(2004)}]{mu04}
\bibinfo{author}{\bibfnamefont{M.}~\bibnamefont{Mukherjee}}
  \bibnamefont{et~al.}, \bibinfo{journal}{Phys.\ Rev.\ Lett.}
  \textbf{\bibinfo{volume}{93}}, \bibinfo{pages}{150801}
  (\bibinfo{year}{2004}).

\bibitem[{\citenamefont{Iacob et~al.}(2006)}]{ia06}
\bibinfo{author}{\bibfnamefont{V.~E.} \bibnamefont{Iacob}}
  \bibnamefont{et~al.}, \bibinfo{journal}{Phys.\ Rev.\ C}
  \textbf{\bibinfo{volume}{74}}, \bibinfo{pages}{15501} (\bibinfo{year}{2006}).

\bibitem[{\citenamefont{Jungmann}(2006)}]{KVI06}
\bibinfo{author}{\bibfnamefont{K.}~\bibnamefont{Jungmann}},
  \bibinfo{journal}{private communication}  (\bibinfo{year}{2006}).

\bibitem[{\citenamefont{Ames et~al.}(1965)\citenamefont{Ames, Phillips, and
  Glickstein}}]{am65}
\bibinfo{author}{\bibfnamefont{O.}~\bibnamefont{Ames}},
  \bibinfo{author}{\bibfnamefont{E.~A.} \bibnamefont{Phillips}},
  \bibnamefont{and} \bibinfo{author}{\bibfnamefont{S.~S.}
  \bibnamefont{Glickstein}}, \bibinfo{journal}{Phys.\ Rev.}
  \textbf{\bibinfo{volume}{137}}, \bibinfo{pages}{B1157}
  (\bibinfo{year}{1965}).

\bibitem[{\citenamefont{Raghavan}(1989)}]{ra89}
\bibinfo{author}{\bibfnamefont{P.}~\bibnamefont{Raghavan}},
  \bibinfo{journal}{At.\ Nucl.\ Data Tables} \textbf{\bibinfo{volume}{42}},
  \bibinfo{pages}{189} (\bibinfo{year}{1989}).

\bibitem[{\citenamefont{Gl$\ddot{\mathrm{u}}$ck}(1997)}]{gl97}
\bibinfo{author}{\bibfnamefont{F.}~\bibnamefont{Gl$\ddot{\mathrm{u}}$ck}},
  \bibinfo{journal}{Comp.\ Phys.\ Comm.} \textbf{\bibinfo{volume}{101}},
  \bibinfo{pages}{223} (\bibinfo{year}{1997}).

\bibitem[{\citenamefont{Scielzo et~al.}(2003)\citenamefont{Scielzo, Freedman,
  Fujikawa, and Vetter}}]{CSD}
\bibinfo{author}{\bibfnamefont{N.~D.} \bibnamefont{Scielzo}},
  \bibinfo{author}{\bibfnamefont{S.~J.} \bibnamefont{Freedman}},
  \bibinfo{author}{\bibfnamefont{B.~K.} \bibnamefont{Fujikawa}},
  \bibnamefont{and} \bibinfo{author}{\bibfnamefont{P.~A.}
  \bibnamefont{Vetter}}, \bibinfo{journal}{Phys.\ Rev.\ A}
  \textbf{\bibinfo{volume}{68}}, \bibinfo{pages}{022716}
  (\bibinfo{year}{2003}).

\bibitem[{\citenamefont{Scielzo
  et~al.}(2004{\natexlab{b}})\citenamefont{Scielzo, Freedman, Fujikawa,
  Kominis, Maruyama, Vetter, and Vieregg}}]{sc04a}
\bibinfo{author}{\bibfnamefont{N.~D.} \bibnamefont{Scielzo}},
  \bibinfo{author}{\bibfnamefont{S.~J.} \bibnamefont{Freedman}},
  \bibinfo{author}{\bibfnamefont{B.~K.} \bibnamefont{Fujikawa}},
  \bibinfo{author}{\bibfnamefont{I.}~\bibnamefont{Kominis}},
  \bibinfo{author}{\bibfnamefont{R.}~\bibnamefont{Maruyama}},
  \bibinfo{author}{\bibfnamefont{P.~A.} \bibnamefont{Vetter}},
  \bibnamefont{and} \bibinfo{author}{\bibfnamefont{J.~R.}
  \bibnamefont{Vieregg}}, \bibinfo{journal}{Nucl. Phys. A}
  \textbf{\bibinfo{volume}{746}}, \bibinfo{pages}{677c}
  (\bibinfo{year}{2004}{\natexlab{b}}).

\bibitem[{\citenamefont{Fatemi et~al.}(2002)\citenamefont{Fatemi, Jones, Lett,
  and Tiesinga}}]{fa02}
\bibinfo{author}{\bibfnamefont{F.~K.} \bibnamefont{Fatemi}},
  \bibinfo{author}{\bibfnamefont{K.~M.} \bibnamefont{Jones}},
  \bibinfo{author}{\bibfnamefont{P.~D.} \bibnamefont{Lett}}, \bibnamefont{and}
  \bibinfo{author}{\bibfnamefont{E.}~\bibnamefont{Tiesinga}},
  \bibinfo{journal}{Phys.\ Rev.\ A} \textbf{\bibinfo{volume}{66}},
  \bibinfo{pages}{053401} (\bibinfo{year}{2002}).

\bibitem[{\citenamefont{Fioretti et~al.}(1998)\citenamefont{Fioretti, Comparat,
  Crubellier, Dulieu, Masnou-Seeuws, and Pillet}}]{fi98}
\bibinfo{author}{\bibfnamefont{A.}~\bibnamefont{Fioretti}},
  \bibinfo{author}{\bibfnamefont{D.}~\bibnamefont{Comparat}},
  \bibinfo{author}{\bibfnamefont{A.}~\bibnamefont{Crubellier}},
  \bibinfo{author}{\bibfnamefont{O.}~\bibnamefont{Dulieu}},
  \bibinfo{author}{\bibfnamefont{F.}~\bibnamefont{Masnou-Seeuws}},
  \bibnamefont{and} \bibinfo{author}{\bibfnamefont{P.}~\bibnamefont{Pillet}},
  \bibinfo{journal}{Phys.\ Rev.\ Lett.} \textbf{\bibinfo{volume}{80}},
  \bibinfo{pages}{4402} (\bibinfo{year}{1998}).

\bibitem[{\citenamefont{Vanhaecke et~al.}(2002)\citenamefont{Vanhaecke,
  de~Souza~Melo, Tolra, Comparat, and Pillet}}]{va02}
\bibinfo{author}{\bibfnamefont{N.}~\bibnamefont{Vanhaecke}},
  \bibinfo{author}{\bibfnamefont{W.}~\bibnamefont{de~Souza~Melo}},
  \bibinfo{author}{\bibfnamefont{B.~L.} \bibnamefont{Tolra}},
  \bibinfo{author}{\bibfnamefont{D.}~\bibnamefont{Comparat}}, \bibnamefont{and}
  \bibinfo{author}{\bibfnamefont{P.}~\bibnamefont{Pillet}},
  \bibinfo{journal}{Phys.\ Rev.\ Lett.} \textbf{\bibinfo{volume}{89}},
  \bibinfo{pages}{063001} (\bibinfo{year}{2002}).

\bibitem[{\citenamefont{Vorobel et~al.}(2003)}]{vo03}
\bibinfo{author}{\bibfnamefont{V.}~\bibnamefont{Vorobel}} \bibnamefont{et~al.},
  \bibinfo{journal}{Eur.\ Phys.\ J.\ A} \textbf{\bibinfo{volume}{16}},
  \bibinfo{pages}{139} (\bibinfo{year}{2003}).

\bibitem[{\citenamefont{Magnier et~al.}(1993)\citenamefont{Magnier, Milli\'e,
  Dulieu, and Masnou-Seeuws}}]{ma93}
\bibinfo{author}{\bibfnamefont{S.}~\bibnamefont{Magnier}},
  \bibinfo{author}{\bibfnamefont{P.}~\bibnamefont{Milli\'e}},
  \bibinfo{author}{\bibfnamefont{O.}~\bibnamefont{Dulieu}}, \bibnamefont{and}
  \bibinfo{author}{\bibfnamefont{F.}~\bibnamefont{Masnou-Seeuws}},
  \bibinfo{journal}{J.\ Chem.\ Phys.} \textbf{\bibinfo{volume}{98}},
  \bibinfo{pages}{7113} (\bibinfo{year}{1993}).

\bibitem[{\citenamefont{Amelink and van~der Straten}(2003)}]{am03}
\bibinfo{author}{\bibfnamefont{A.}~\bibnamefont{Amelink}} \bibnamefont{and}
  \bibinfo{author}{\bibfnamefont{P.}~\bibnamefont{van~der Straten}},
  \bibinfo{journal}{Phys.\ Scripta} \textbf{\bibinfo{volume}{68}},
  \bibinfo{pages}{C82} (\bibinfo{year}{2003}).

\bibitem[{\citenamefont{Luhmann}(1997)}]{lu97}
\bibinfo{author}{\bibfnamefont{T.}~\bibnamefont{Luhmann}},
  \bibinfo{journal}{Rev.\ Sci.\ Instrum.} \textbf{\bibinfo{volume}{68}},
  \bibinfo{pages}{2347} (\bibinfo{year}{1997}).

\bibitem[{\citenamefont{Ketterle et~al.}(1993)\citenamefont{Ketterle, Davis,
  Joffe, Martin, and Pritchard}}]{ke93}
\bibinfo{author}{\bibfnamefont{W.}~\bibnamefont{Ketterle}},
  \bibinfo{author}{\bibfnamefont{K.~B.} \bibnamefont{Davis}},
  \bibinfo{author}{\bibfnamefont{M.~A.} \bibnamefont{Joffe}},
  \bibinfo{author}{\bibfnamefont{A.}~\bibnamefont{Martin}}, \bibnamefont{and}
  \bibinfo{author}{\bibfnamefont{D.~E.} \bibnamefont{Pritchard}},
  \bibinfo{journal}{Phys.\ Rev.\ Lett.} \textbf{\bibinfo{volume}{70}},
  \bibinfo{pages}{2253} (\bibinfo{year}{1993}).

\bibitem[{\citenamefont{Hardy and Towner}(2005)}]{ha05}
\bibinfo{author}{\bibfnamefont{J.~C.} \bibnamefont{Hardy}} \bibnamefont{and}
  \bibinfo{author}{\bibfnamefont{I.~S.} \bibnamefont{Towner}},
  \bibinfo{journal}{Phys.\ Rev.\ C} \textbf{\bibinfo{volume}{71}},
  \bibinfo{pages}{055501} (\bibinfo{year}{2005}).

\bibitem[{\citenamefont{Carlson}(1963)}]{al63}
\bibinfo{author}{\bibfnamefont{T.}~\bibnamefont{Carlson}},
  \bibinfo{journal}{Phys.\ Rev.} \textbf{\bibinfo{volume}{132}},
  \bibinfo{pages}{2239} (\bibinfo{year}{1963}).

\bibitem[{\citenamefont{Johnson et~al.}(1963)\citenamefont{Johnson, Pleasonton,
  and Carlson}}]{jo63}
\bibinfo{author}{\bibfnamefont{C.~H.} \bibnamefont{Johnson}},
  \bibinfo{author}{\bibfnamefont{F.}~\bibnamefont{Pleasonton}},
  \bibnamefont{and} \bibinfo{author}{\bibfnamefont{T.~A.}
  \bibnamefont{Carlson}}, \bibinfo{journal}{Phys.\ Rev.}
  \textbf{\bibinfo{volume}{132}}, \bibinfo{pages}{1149} (\bibinfo{year}{1963}).

\bibitem[{\citenamefont{Stratowa et~al.}(1978)\citenamefont{Stratowa,
  Dobrozemsky, and Weinzierl}}]{st78}
\bibinfo{author}{\bibfnamefont{C.}~\bibnamefont{Stratowa}},
  \bibinfo{author}{\bibfnamefont{R.}~\bibnamefont{Dobrozemsky}},
  \bibnamefont{and}
  \bibinfo{author}{\bibfnamefont{P.}~\bibnamefont{Weinzierl}},
  \bibinfo{journal}{Phys.\ Rev.\ D} \textbf{\bibinfo{volume}{18}},
  \bibinfo{pages}{3970} (\bibinfo{year}{1978}).

\bibitem[{\citenamefont{Byrne et~al.}(2002)\citenamefont{Byrne, Dawber, van~der
  Grinten, Habeck, Shaikh, Spain, Scott, Baker, Green, and Zimmer}}]{by02}
\bibinfo{author}{\bibfnamefont{J.}~\bibnamefont{Byrne}},
  \bibinfo{author}{\bibfnamefont{P.~G.} \bibnamefont{Dawber}},
  \bibinfo{author}{\bibfnamefont{M.~G.~D.} \bibnamefont{van~der Grinten}},
  \bibinfo{author}{\bibfnamefont{C.~G.} \bibnamefont{Habeck}},
  \bibinfo{author}{\bibfnamefont{F.}~\bibnamefont{Shaikh}},
  \bibinfo{author}{\bibfnamefont{J.~A.} \bibnamefont{Spain}},
  \bibinfo{author}{\bibfnamefont{R.~D.} \bibnamefont{Scott}},
  \bibinfo{author}{\bibfnamefont{C.~A.} \bibnamefont{Baker}},
  \bibinfo{author}{\bibfnamefont{K.}~\bibnamefont{Green}}, \bibnamefont{and}
  \bibinfo{author}{\bibfnamefont{O.}~\bibnamefont{Zimmer}},
  \bibinfo{journal}{J.\ Phys.\ G} \textbf{\bibinfo{volume}{28}},
  \bibinfo{pages}{1325} (\bibinfo{year}{2002}).

\bibitem[{\citenamefont{Adelberger et~al.}(1999)}]{ad99}
\bibinfo{author}{\bibfnamefont{E.~G.} \bibnamefont{Adelberger}}
  \bibnamefont{et~al.}, \bibinfo{journal}{Phys.\ Rev.\ Lett.}
  \textbf{\bibinfo{volume}{83}}, \bibinfo{pages}{1299} (\bibinfo{year}{1999}).

\bibitem[{\citenamefont{Egorov et~al.}(1997)}]{eg97}
\bibinfo{author}{\bibfnamefont{V.}~\bibnamefont{Egorov}} \bibnamefont{et~al.},
  \bibinfo{journal}{Nucl.\ Phys.\ A} \textbf{\bibinfo{volume}{621}},
  \bibinfo{pages}{745} (\bibinfo{year}{1997}).

\end{thebibliography}

\end{document}